\documentclass[aps,pra,twocolumn,showpacs,superscriptaddress,floatfix,nofootinbib]{revtex4}
\usepackage{graphicx}
\usepackage{nicefrac}
\usepackage{amsmath}
\usepackage{amsfonts}
\usepackage{amssymb}
\usepackage{amsthm}
\usepackage{epsf}
\usepackage{bm}
\usepackage{bbm}
\usepackage{longtable}

\usepackage{dcolumn}

\newcolumntype{.}{D{x}{}{-1}}

\newcommand{\balpha}{\bm{\alpha}}

\newcommand{\bfr}{\bm{r}}
\newcommand{\bfk}{\bm{k}}
\newcommand{\bfp}{\bm{p}}

\newcommand{\hp}{\hat{\bfp}}

\newcommand{\vare}{\varepsilon}

\newcommand{\SixJ}[6]{
        \left\{
        \begin{array}{ccc}
        #1  & #2  & #3 \\
        #4  & #5  & #6 \\
        \end{array}
        \right\}
        }
\newcommand{\NineJ}[9]{
        \left\{
          \begin{array}{ccc}
            #1  & #2  & #3 \\
            #4  & #5  & #6 \\
            #7  & #8  & #9 \\
          \end{array}
        \right\} }

\newcommand{\lbr}{\left<} \newcommand{\rbr}{\right>}

\begin{document}

\title{Comparative study of the electron- and positron-atom bremsstrahlung}

\author{V. A. Yerokhin}
\affiliation{Institute of Physics, Heidelberg University,
Im Neuenheimer Feld 226, D-69120 Heidelberg, Germany}
\affiliation{GSI Helmholtzzentrum f\"ur Schwerionenforschung, Planckstra{\ss}e 1,
D-64291 Darmstadt, Germany}
\affiliation{Center for Advanced Studies, St.~Petersburg State
Polytechnical University, Polytekhnicheskaya 29,
St.~Petersburg 195251, Russia}

\author{A. Surzhykov}
\affiliation{Institute of Physics, Heidelberg University,
Im Neuenheimer Feld 226, D-69120 Heidelberg, Germany}
\affiliation{GSI Helmholtzzentrum f\"ur Schwerionenforschung, Planckstra{\ss}e 1,
D-64291 Darmstadt, Germany}

\author{R. M\"artin}
\affiliation{Helmholtz-Institut Jena, Fr\"obelstieg 3, D-07743 Jena, Germany}
\affiliation{GSI Helmholtzzentrum f\"ur Schwerionenforschung, Planckstra{\ss}e 1,
D-64291 Darmstadt, Germany}

\author{S. Tashenov}
\affiliation{Institute of Physics, Heidelberg University,
Im Neuenheimer Feld 226, D-69120 Heidelberg, Germany}

\author{G. Weber}
\affiliation{Helmholtz-Institut Jena, Fr\"obelstieg 3, D-07743 Jena, Germany}
\affiliation{GSI Helmholtzzentrum f\"ur Schwerionenforschung, Planckstra{\ss}e 1,
D-64291 Darmstadt, Germany}

\begin{abstract}

Fully relativistic treatment of the electron-atom and positron-atom
bremsstrahlung is reported. The calculation is based on the
partial-wave expansion of the Dirac scattering states in 
an external atomic field. A comparison of the electron and positron 
bremsstrahlung is presented for
the single and double differential cross sections and the 
Stokes parameters of the emitted photon. It is demonstrated that the
electron-positron symmetry of the bremsstrahlung spectra, which is nearly exact in the
nonrelativistic regime, is to a large extent removed by the relativistic
effects. 

\end{abstract}


\maketitle

When a charged particle traverses an atomic field, a part of its energy may be
converted into radiation. This is the atomic bremsstrahlung, one of the
fundamental collision processes. It is one of the important mechanisms of the
energy loss in hot plasmas and in particle beams traversing thick 
targets. 
The process of the {\em electron-atom} bremsstrahlung has been extensively studied in the
literature. The single and double differential cross sections of this process were
tabulated \cite{pratt:77,kissel:83} and the polarization properties of the emitted
radiation were calculated 
\cite{tseng:71,tseng:73,yerokhin:10:bs,jakubassa:10}. 

Much less is known on the {\em positron-atom} bremsstrahlung. Several
reported investigations of this process \cite{jabbur:63,feng:81,kim:86}
were dealing primarily with the bremsstrahlung energy loss. It was shown that,
for high energies of the incident and final projectile, the 
cross sections of the electron and positron bremsstrahlung are
very much similar. 
However, when the energy of the final projectile decreases, the cross section
of the positron bremsstrahlung becomes increasingly suppressed as compared to the electron
one, because of the strong Coulomb repulsion between the positron and nucleus 
at short distances. The ratio of the positron-to-electron bremsstrahlung
stopping power was the main subject of those early works. 

To the best of our knowledge, the angular dependence of the cross section and the
polarization properties of the relativistic positron-atom bremsstrahlung 
have never been studied. They are becoming subjects of experimental interest 
today, with the advent of techniques for the production of highly polarized
positrons beams \cite{alexander:08} and the detection of polarization
correlations in the bremsstrahlung radiation \cite{tashenov:11,maertin:12}. 

In the present investigation, we make a comparative study of the positron- and
electron-atom bremsstrahlung, by analyzing the double
differential cross section and the polarization correlations between the
incident projectile and the emitted photon. The calculation is performed 
within the fully relativistic approach based on the
partial-wave representation of the Dirac continuum states with a
fixed value of the asymptotic momentum. This work extends our previous
calculations of the electron-atom bremsstrahlung 
\cite{yerokhin:10:bs,weber:12}.

\section{Theory}

Relativistic theory of the electron-atom bremsstrahlung is nowadays well
established and can be found, e.g., in the review article \cite{pratt:84:book}. 
For our bremsstrahlung calculations we use the density-matrix formalism described
in detail in our previous paper \cite{yerokhin:10:bs}. In this section, we will
extend the formulas obtained previously
to the case of the positron bremsstrahlung. Such extension can be done by
exploiting the fact that in the QED
theory,
an incoming positron
with a four-momentum $p_i$ and a helicity $m_i$ can be described as an outgoing
electron with a four-momentum $-p_i$ and a helicity $-m_i$  (see, e.g., books \cite{bjorken:65,itzykson:80}).

We now consider the transition from the electron-atom to the positron-atom
bremsstrahlung in more detail. The amplitude of the electron-atom
bremsstrahlung is
\begin{align} \label{1}
M_{if}^{\rm el}(\lambda) = &\ \int d\bm{r}\,
\Psi^{(+)^{\dag}}(\vare_i,\bm{p}_i,m_i)\,
  \nonumber \\ & \times
\balpha \cdot \hat{\bm{u}}_{\lambda}\,
e^{i\bfk\cdot\bfr}\, \Psi^{(-)}(\vare_f,\bm{p}_f,m_f)\,,
\end{align}
where $\Psi^{(+)}(\vare_i,\bm{p}_i,m_i)$ is the wave function of the inital
electron state with the energy $\vare_i$, the momentum $\bm{p}_i$, the helicity
$m_i$, and the asymptotics of an outgoing spherical wave, and 
$\Psi^{(-)}(\vare_f,\bm{p}_f,m_f)$ is the wave function of the
final electron state with the energy $\vare_f$, the momentum $\bm{p}_f$, the helicity
$m_f$, and the asymptotics of an incoming spherical wave,
and $\bfk$ and $\lambda$ are the momentum and polarization of the emitted photon, respectively. 
The Dirac scattering states $\Psi^{(\pm)}(\vare,\bm{p},m)$ are given by
their partial-wave expansion \cite{rose:61,eichler:95:book}
\begin{align} \label{2}
\Psi^{(\pm)}(\vare,\bm{p},m)  = \frac1{\sqrt{p\,|\vare|}}\, \sum_{\kappa\mu} i^l\, e^{\pm i\Delta_{\kappa}}\,C^{j\mu}_{l m_l, \frac12
  m}\, Y_{lm_l}^*(\hp)\,   |\vare \kappa \mu\bigr>
\,,
\end{align}
where $|\vare \kappa \mu\bigr>$ are the Dirac continuum states with a given
angular-momentum quantum number $\kappa$ and angular-momentum projection
$\mu$, $j = |\kappa|-\nicefrac12$, $l = |\kappa+\nicefrac12|-\nicefrac12$, 
and $\Delta_{\kappa} =
\sigma_{\kappa} - \sigma^{(0)}_{\kappa}$ is the difference between the
asymptotic large-distance phase of the Dirac-Coulomb solution and the free Dirac solution
(see book \cite{rose:61} for details). 

In order to obtain the amplitude of the positron-atom bremsstrahlung from
Eq.~(\ref{1}), we need to make the following substitutions in the wave
functions: $\vare\to-\vare$, $\bm{p}\to-\bm{p}$, $m\to -m$, 
$\Psi^{(\pm)}\to \Psi^{(\mp)}$, and to interchange the initial and the final
state. The resulting 
amplitude of the positron-atom bremsstrahlung is given by
\begin{align} \label{3}
M_{if}^{\rm pos}(\lambda) = &\ \int d\bm{r}\,
\Psi^{(+)^{\dag}}(-\vare_f,-\bm{p}_f,-m_f)\,
  \nonumber \\ & \times
\balpha\cdot\hat{\bm{u}}_{\lambda}\,
e^{i\bfk\cdot\bfr}\, \Psi^{(-)}(-\vare_i,-\bm{p}_i,-m_i)\,,
\end{align}
where $\bfp_i(\bfp_f)$ and $m_i(m_f)$ are the momentum and helicity of the
initial (final) state positron, respectively.

Assuming that the final-state positron is not observed in the experiment,
we introduce
the two-by-two reduced density matrix of the final (photon) state, 
\begin{align} \label{e1}
\lbr \bfk \lambda| \rho_f | \bfk \lambda^{\prime}\rbr
 = &\ \sum_{m_i m_{i'} m_f} \int d\Omega_f\, M_{if}^*(\lambda)\,
 M_{i'f}(\lambda')\,
  \nonumber \\ & \times
 \lbr \bfp_im_i|\rho_i|\bfp_im_{i'}\rbr\,,
\end{align}
where 
$\Omega_f$ is the solid angle of the scattered positron, and 
$\lbr \bfp_im_i|\rho_i|\bfp_im_{i'}\rbr$ is the density matrix of the
initial positron state. Note that the inital density matrix is the same 
for positrons and electrons (since it depends only on the spin of the particle and
not on its charge). The photon direction $\hat{\bfk} = \bfk/k$ will be
characterized by the Euler angles $(\theta,\phi)$, with the $z$ axis directed
along the momentum of the initial-state projectile $\bfp_i$. The final-state
density matrix (\ref{e1}) contains all the information needed
for calculating the differential cross section and the polarization
correlations of the bremsstrahlung radiation. 

Extending the derivation given in Ref.~\cite{yerokhin:10:bs} 
to the case of the positron-atom bremsstrahlung, we obtain
\begin{widetext}
\begin{align} \label{14}
\bigl< \bfk \lambda| {\rho}_f |\bfk \lambda^{\prime}\bigr>
 =  &\ 8(2\pi)^4\, \sum_{\kappa_i \kappa_i^{\prime}
    \kappa_f} \sum_{LL^{\prime}\kappa g t}
  \sum_{\gamma_1 \gamma_2}
D^{g^*}_{\gamma_1\gamma_2}(\hat{\bfk})\,
  (-1)^{\kappa}\rho^{(i)}_{\kappa,\gamma_1}\, 
  i^{l_i-l_i^{\prime}-L+L^{\prime}}\,
  e^{i\Delta_{\kappa_i,-\vare_i}-i\Delta_{\kappa_i^{\prime},-\vare_i}}\,
 \left[L,L^{\prime},j_i,j_i^{\prime},l_i,l_i^{\prime},g,\kappa\right]^{1/2}\,
\nonumber \\ &\times
(-1)^{j_i^{\prime}-j_f+l_i+g+\kappa}\,
   C_{L^{\prime}-\lambda^{\prime}, L\lambda}^{g\gamma_2}\,
   C_{l_i0,l_i^{\prime}0}^{t0}\,C_{g-\gamma_1,\kappa\gamma_1}^{t0}\,
  \SixJ{L}{j_f}{j_i}{j_i^{\prime}}{g}{L^{\prime}}\,
  \NineJ{\nicefrac12}{\nicefrac12}{\kappa}{j_i^{\prime}}{j_i}{g}{l_i^{\prime}}{l_i}{t}
\nonumber \\ &\times
 \sum_{pp^{\prime}}
    (i\lambda)^p\, (-i\lambda^{\prime})^{p^{\prime}}\,
 \left< -\vare_i\kappa_i\left|\left| \balpha\cdot \bm{a}^{(p)}_{L}\right|\right|-\vare_f\kappa_f\right>\,
 \left< -\vare_i\kappa_i^{\prime}\left|\left| \balpha\cdot
 \bm{a}^{(p^{\prime})}_{L^{\prime}}
               \right|\right|-\vare_f\kappa_f\right>^*\,,
\end{align}
\end{widetext}
where $D^L_{M\lambda}$ is Wigner's $D$ function \cite{varshalovich},
$\rho^{(i)}_{\kappa}$ is the spherical tensor of the initial-state density
matrix [see Eq.~(10) of Ref.~\cite{yerokhin:10:bs}], 
${\bm a}^{(p)}_L$ are the magnetic ($p = 0$) and electric ($p = 1$) operators defined
by Eqs.~(15)-(17) of Ref.~\cite{yerokhin:10:bs},
$[x_1,x_2,\ldots] \equiv (2x_1+1)(2x_2+1)\ldots$. 
The states $|-\vare_i\kappa_i\rangle$ and
$|-\vare_f\kappa_f\rangle$ are the spherical-wave continuum-state Dirac wave
functions with a {\em negative} energy, $-\vare_i < -m$ and
$-\vare_f < -m$, where $m$ is the electron rest mass. 

In our calculations of the final-state density matrix, we used the method described in
Ref.~\cite{yerokhin:10:bs}, with radial integrals evaluted numerically and
the integration contour rotated to the imaginary axis. 
We performed calculations for two types of the target, (i)
the bare nucleus and 
(ii) the neutral atom. In the latter case, the electronic structure of the atom was
represented by a static screening potential obtained by the Dirac-Fock
method. 

The negative-energy continuum-state Dirac wave functions
$|-\vare\kappa\rangle$ for the point Coulomb potential can be calculated by
using their analytic representation in terms of the
Whittaker $M$ function \cite{eichler:95:book}, similarly to that
for the positive energies. For the neutral atoms, however, the wave functions have
to be computed numerically. In this case, it is convenient to transform the
negative-energy states to the positive-energy ones by using the symmetry of the
radial Dirac equation. One can show that the upper and lower radial
components of the negative-energy Dirac solution with the potential $V$,
$g_{-\vare,\kappa}^{V}$  and $f_{-\vare,\kappa}^{V}$, can be expressed in
terms of the components of the positive-energy solutions with the potential $-V$
as
\begin{align}
g_{-\vare,\kappa}^{V}(r) = f_{\vare,-\kappa}^{-V}(r)\,,\\
f_{-\vare,\kappa}^{V}(r) = g_{\vare,-\kappa}^{-V}(r)\,.
\end{align}
This approach to the evaluation of the negative-energy Dirac states 
was previously used in Ref.~\cite{artemyev:03:ncdr}.

\section{Results and discussion}

We begin with calculating the single-differential cross section of the
electron and positron-atom bremsstrahlung. The results can be conveniently
represented in terms of the scaled cross section
$\sigma \equiv  (k/Z^2)\, d\sigma/dk$, where $Z$ is the nuclear charge and $k$
is the photon energy. This part of our calculations can be compared with the
previous work by Feng, Pratt, and Tseng \cite{feng:81}. Very good agreement 
with their calculation was found. So, for the inital
kinetic energy $E = 50$~keV, the fractional energy carried by the photon
$k/E = 0.6$, and the bare-nucleus target,
we obtain the ratio of the positron-to-electron bremsstrahlung cross sections 
$\sigma^+/\sigma^- = 0.61498$ for $Z=8$ and $0.003839$
for $Z=92$, whereas Ref.~\cite{feng:81} reports 
$0.615$ and $0.00384$, respectively.

\begin{figure*}
  \centerline{\includegraphics[width=0.9\textwidth]{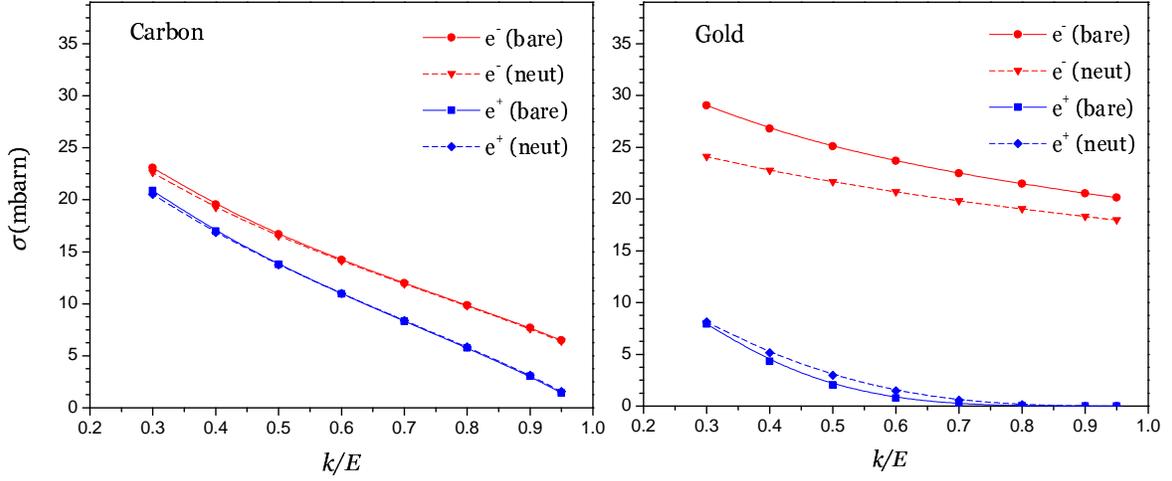}}
\caption{(Color online) Single-differential cross section $\sigma \equiv
  (k/Z^2)\, d\sigma/dk$ for the initially unpolarized electrons
  ($e^-$, red) and positrons ($e^+$, blue), for the 
  carbon (left) and gold (right)
  targets, as a function of the fractional energy carried by the photon. 
  The initial energy of projectile is fixed by $E = 100$~keV.
  Calculational results for the bare nucleus are shown by solid line and those
  for the neutral atom, by the dashed line.
  \label{fig:tot} }
\end{figure*}

Our numerical results for the cross sections of the positron and electron 
bremsstrahlung are presented in Fig.~\ref{fig:tot} for two
targets, carbon and gold, and the initial projectile energy of 100~keV. 
Carbon is an essentially nonrelativistic system. 
In this case, the positron bremsstrahlung is only slightly suppressed
as compared to the electron one and the difference between the bare
nucleus and neutral atom is very small (i.e., the screening effect of the
atomic electrons is nearly negligible).  For gold, on the contrary, the 
relativistic binding effects are large. Within the classical-physics picture,
one can expect that the strong Coulomb potential of the nucleus 
significantly changes the projectile velocity at the point of the closest 
approach (where the photon emission is most probable) and thus breaks the
symmetry between the electron and the positron spectra.
Indeed, our calculations for gold show a large suppression of the positron
bremsstrahlung in the region $k/E > 0.5$ and also a significant 
screening effect of the target electrons. 
We observe that the screening effect reduces the cross section for the case of the
electron bremsstrahlung but enhances it for the positron bremsstrahlung.
Remarkably, in the region $k/E \sim 1$,
the screening effect enhances the positron bremsstrahlung by an order of
magnitude as compared to the pure Coulomb field (as noted already in 
Ref.~\cite{feng:81}). But, as the resulting cross section is still very small, 
the effect is difficult to observe experimentally.

\begin{figure*}
  \centerline{\includegraphics[width=\textwidth]{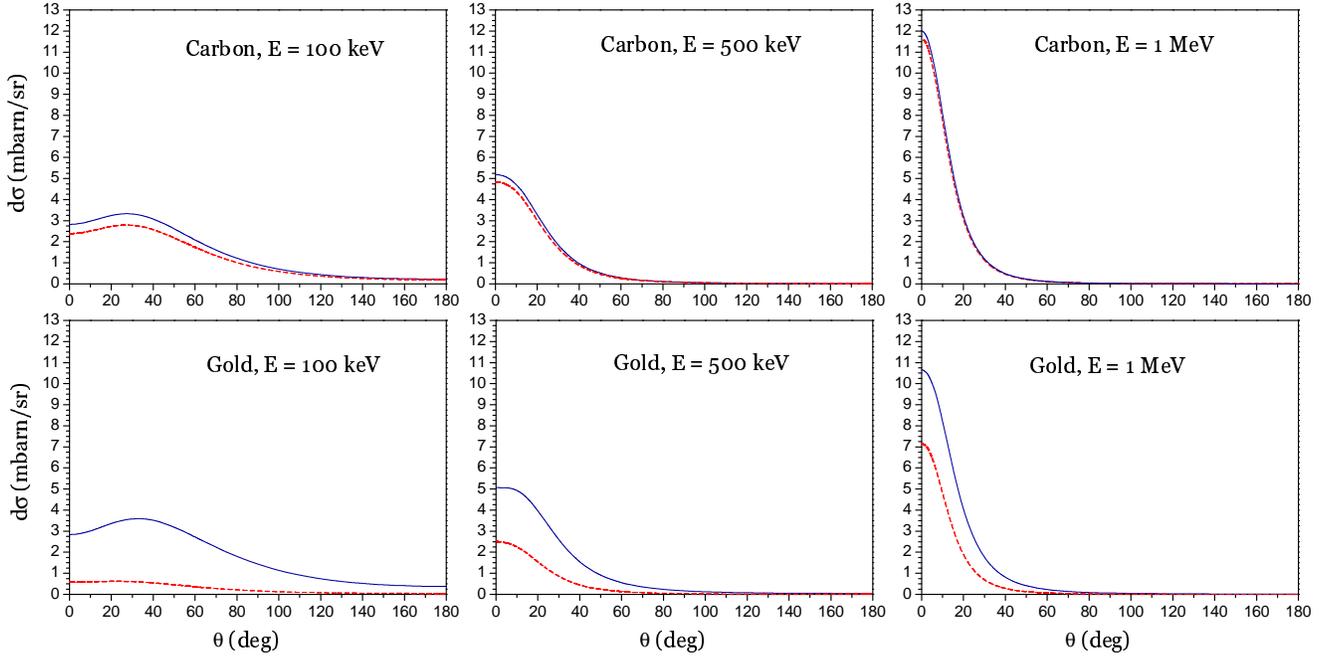}}
\caption{(Color online) Double-differential cross section $d\sigma \equiv
  (k/Z^2)\, d\sigma/(dk\,d\Omega_k)$ for the initially unpolarized electrons
  (blue, solid line) and positrons (red, dashed line), 
  for the carbon (top) and gold (bottom)
  targets, as a function of the photon emission angle, for different initial
  energies of the projectile. 
  The fractional energy carried by photon is $k/E = 0.5$. 
  The calculation was performed for neutral atomic targets.
  \label{fig:diff} }
\end{figure*}

The numerical results for the double-differential cross section, $d\sigma 
\equiv  (k/Z^2)\, d\sigma/(dk\,d\Omega_k)$, 
are presented in Fig.~\ref{fig:diff}. 
The calculation was performed for neutral atomic targets (carbon and gold) and
several energies of the incoming projectile.
We observe that the dominant difference
between the electron $d\sigma^-$ and positron $d\sigma^+$ cross sections
comes from the suppression of the positron bremsstrahlung. As can be
seen from the picture, the suppression grows with increasing the nuclear charge
and decreasing the energy of the initial projectile. This is in agreement with 
Ref.~\cite{kim:86}, which concludes that, for a large range of 
the kinetic energy of the incoming projectile $E_{\rm kin}$, 
the overall suppression factor is a nearly linear function of $Z^2/E_{\rm kin}$.

\begin{figure*}
  \centerline{\includegraphics[width=\textwidth]{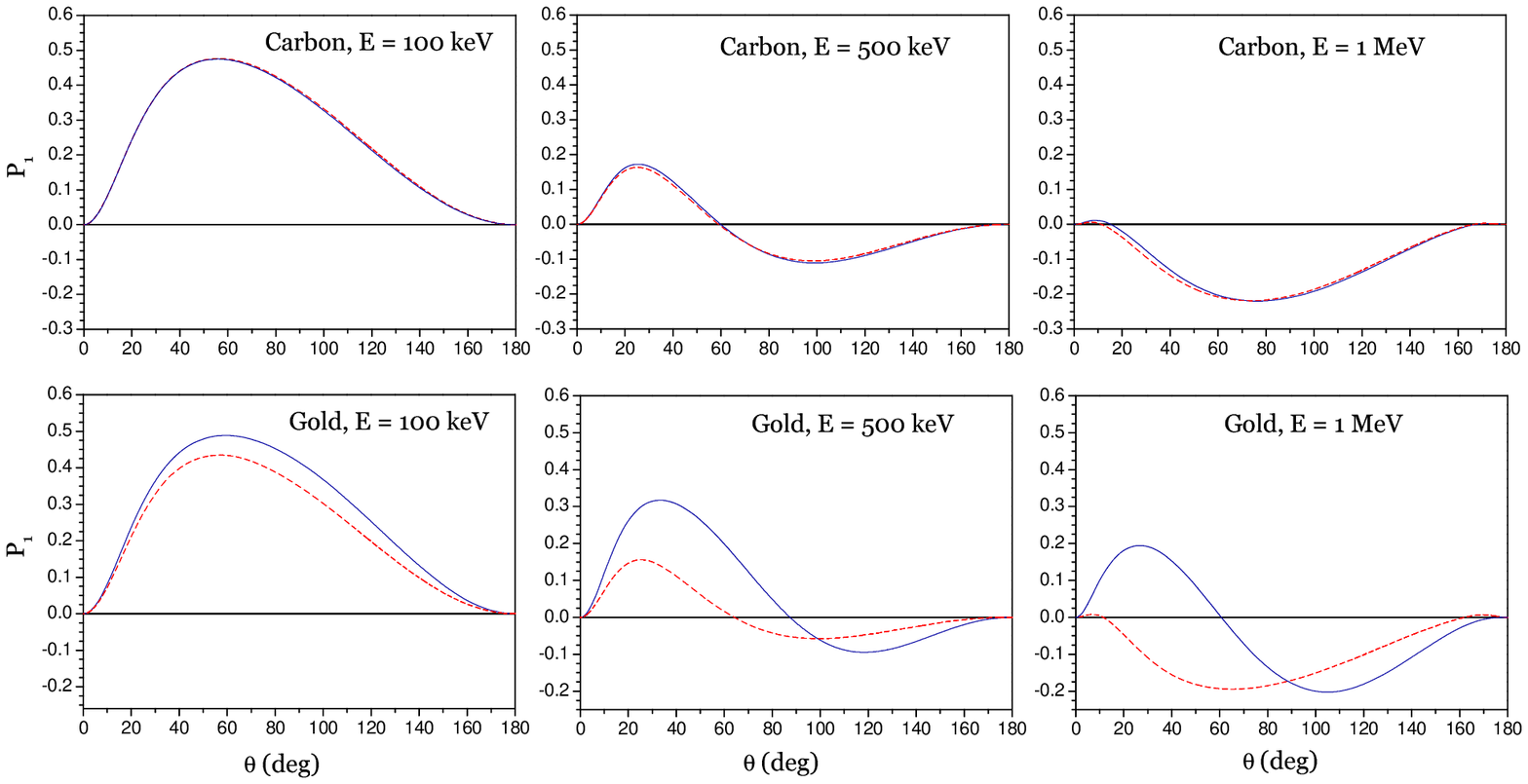}}
\caption{(Color online) Stokes parameter $P_1$ for the initially unpolarized electrons
  (blue, solid line) and positrons (red, dashed line), 
  for the carbon (top) and gold (bottom)
  targets, as a function of the photon emission angle, for different initial
  energies of the projectile $E$.
  The fractional energy carried by photon is $k/E = 0.5$. 
  \label{fig:P1} }
\end{figure*}

\begin{figure*}
  \centerline{\includegraphics[width=\textwidth]{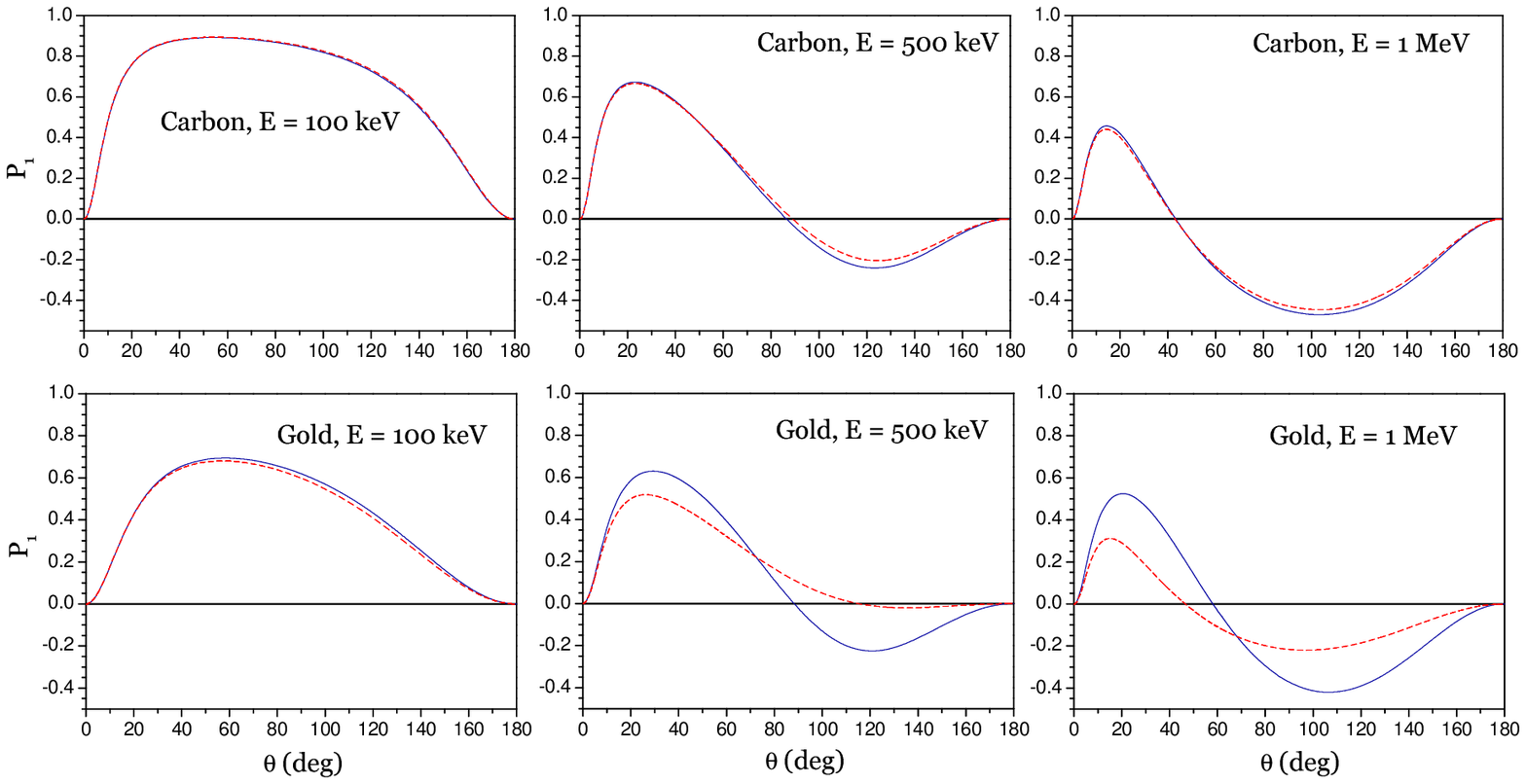}}
\caption{(Color online) Stokes parameter $P_1$ for the initially unpolarized electrons
  (blue, solid line) and positrons (red, dashed line), 
  for the carbon (top) and gold (bottom)
  targets, as a function of the photon emission angle, for different initial
  energies of the projectile $E$.
  The fractional energy carried by photon is $k/E = 0.9$. 
  \label{fig:P1_} }
\end{figure*}

Next, we compare the polarization properties of the electron and positron
bremsstrahlung. The most important polarization property is the 
Stokes parameter of the emitted photon $P_1$ for the initially unpolarized
projectile. (In this case, $P_1$ yields also the degree of the linear
polarization of the emitted radiation.)
The numerical results for $P_1$ are presented in
Figs.~\ref{fig:P1} and \ref{fig:P1_}. 
Since the screening effect of atomic electrons on the Stokes parameters is
rather weak, the results shown were obtained for the bare nucleus targets. 
We again observe that for the light targets (carbon),
the polarization of the positron
bremsstrahlung is almost identical to the electron one. For the
gold target, however, the relativistic effects break
the electron-positron symmetry of the bremsstrahlung radiation.
The difference between the electron and positron bremsstrahlung spectra
becomes increasingly more pronounced when
the initial projectile energy is enlarged. 

\begin{figure*}
  \centerline{\includegraphics[width=\textwidth]{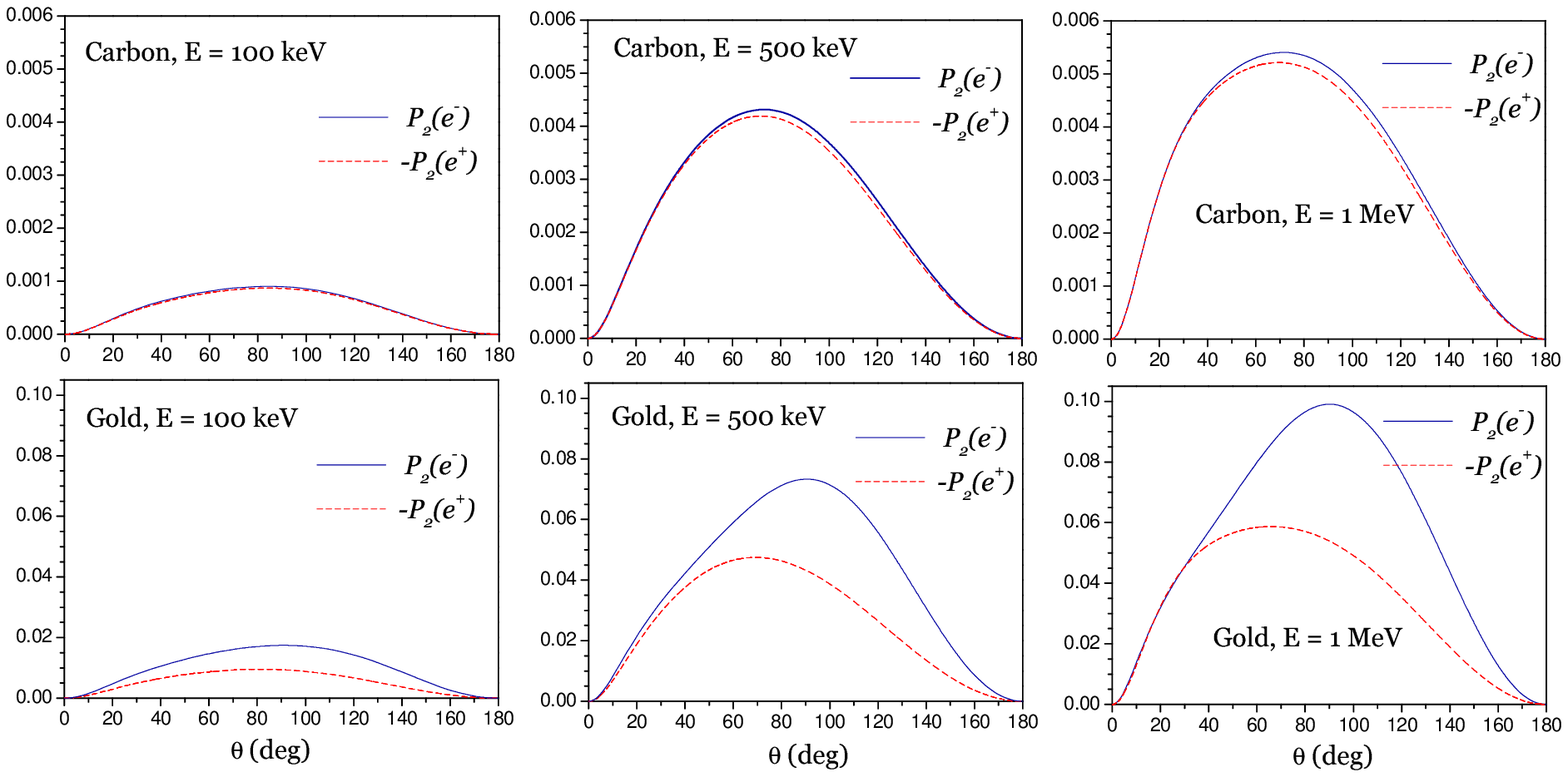}}
\caption{(Color online) Stokes parameter $P_2$ for the initially
  logitudinally polarized electrons
  ($P_2(e^-)$, blue, solid line) and positrons ($P_2(e^+)$, red, dashed line), 
  for the carbon (top) and gold (bottom)
  targets, as a function of the photon emission angle, for different initial
  energies of the projectile $E$.
  The fractional energy carried by photon is $k/E = 0.5$. 
  Note that, for positrons, $P_2$ with the reversed overall sign is plotted. 
  \label{fig:P2} }
\end{figure*}

The second polarization correlation that is of the experimental interest
today is the Stokes parameter of the emitted photon $P_2$. 
It manifests itself as a rotation
of the polarization ellipse of the emitted radiation
in the plane perpendicular to the photon momentum.
The ratio of $P_2$ and $P_1$ yields the
tilt angle $\chi$ of the polarization ellipse ($P_2/P_1 = \tan 2\chi$), 
which can nowadays be measured very
precisely \cite{tashenov:11}. Our numerical results for the Stokes parameter
$P_2$ for the initially {\em longitudinally} polarized projectile
are presented in Fig.~\ref{fig:P2}. 
Since $P_2$ is purely a relativistic effect, its typical numerical values are very 
small for low-$Z$ target but grow rapidly when the initial projectile energy
and the nuclear charge are increased. 

An important observation is that the numerical values of $P_2$
for the initially longitudinally polarized electrons and positrons are of 
the opposite sign. This could have been anticipated from the fact that
$P_2$ scales linearly with $Z$ \cite{tseng:73} and, therefore, changes
its sign after the substitution $Z\to-Z$. The consequence of the 
opposite sign of $P_2(e^{-})$ and $P_2(e^+)$ is that the rotation of the
polarization ellipse of the emitted radiation for the longitudinally polarized
electron and positron beams will occur in the opposite directions. 

Beside this effect, we observe that
$P_2(e^-)$ and $-P_2(e^+)$ nearly coincide for light targets but become
increasingly different as the nuclear charge and the initial projectile energy
are enlarged. We also note that the difference between the 
Stokes parameters is generally smaller at the forward angles and larger at the
backward angles. This effect can be understood from simple classical-physics
arguments. In order to scatter backwards off the positively-charge nucleus,
the electron has to go all the way around the nucleus, whereas the positron
just enters the nuclear field and bounces back. So, for the back scattering, 
the paths within the region of the strong field (and, therefore, the
accumulated relativistic effects) are very different for the 
electron and the positron. For the forward scattering, however, mainly the
straight trajectories contribute and so the paths of the electron and the
positron in the strong Coulomb field are almost the same.

In summary, we perfomed a calculation of the electron-atom and positron-atom
bremsstrahlung within the rigorous relativistic approach based on the
partial-wave expansion of the Dirac wave functions in the external atomic field.
Comparison between the electron and positron bremsstrahlung spectra is made for
the single and double differential cross sections and the Stokes parameters of the
emitted radiation. It is demonstrated that for the low-$Z$ targets, 
the polarization of the electron and positron bremsstrahlung radiation
is very much similar (except for the polarization correlations vanishing in the
nonrelativistic limit). However, for heavy relativistic targets and high
impact energies, the positron bremsstrahlung becomes significantly 
suppressed and distorted as compared to the electron bremsstrahlung. 

The work reported in this paper was supported by the Helmholtz Gemeinschaft
(Nachwuchsgruppe VH-NG-421). S.T. acknowledges 
the support by the German Research Foundation (DFG) within 
the Emmy Noether program under contract No. TA 740 1-1.

%
%

%
%

%
%

%
%

%
%



\begin{thebibliography}{10}

\bibitem{pratt:77}
R.~H. Pratt, H.~K. Tseng, C.~M. Lee, and L.~Kissel,
\newblock At. Data Nucl. Data Tables {\bf 20}, 175  (1977).

\bibitem{kissel:83}
L.~Kissel, C.~A. Quarles, and R.~H. Pratt,
\newblock At. Data Nucl. Data Tables {\bf 28}, 381  (1983).

\bibitem{tseng:71}
H.~K. Tseng and R.~H. Pratt,
\newblock Phys. Rev. A {\bf 3}, 100 (1971).

\bibitem{tseng:73}
H.~K. Tseng and R.~H. Pratt,
\newblock Phys. Rev. A {\bf 7}, 1502 (1973).

\bibitem{yerokhin:10:bs}
V.~A. Yerokhin and A.~Surzhykov,
\newblock Phys. Rev. A {\bf 82}, 062702 (2010).

\bibitem{jakubassa:10}
D.~H. Jakubassa-Amundsen,
\newblock Phys. Rev. A {\bf 82}, 042714 (2010).

\bibitem{jabbur:63}
R.~J. Jabbur and R.~H. Pratt,
\newblock Phys. Rev. {\bf 129}, 184 (1963).

\bibitem{feng:81}
I.~J. Feng, R.~H. Pratt, and H.~K. Tseng,
\newblock Phys. Rev. A {\bf 24}, 1358 (1981).

\bibitem{kim:86}
L.~Kim, R.~H. Pratt, S.~M. Seltzer, and M.~J. Berger,
\newblock Phys. Rev. A {\bf 33}, 3002 (1986).

\bibitem{alexander:08}
G.~Alexander et~al.,
\newblock Phys. Rev. Lett. {\bf 100}, 210801 (2008).

\bibitem{tashenov:11}
S.~Tashenov, T.~B\"ack, R.~Barday, B.~Cederwall, J.~Enders, A.~Khaplanov,
  Y.~Poltoratska, K.-U. Sch\"assburger, and A.~Surzhykov,
\newblock Phys. Rev. Lett. {\bf 107}, 173201 (2011).

\bibitem{maertin:12}
R.~M\"artin {et al.},
\newblock Phys. Rev. Lett. {\bf 108}, 264801 (2012).

\bibitem{weber:12}
G. Weber, R. M\"artin, A. Surzhykov, M. Yasuda, V. A. Yerokhin, and Th. St\"ohlker,
\newblock Nucl. Instr. Meth. Res. B {\bf 279}, 155 (2012).

\bibitem{pratt:84:book}
R.~H. Pratt and I.~J. Feng,
\newblock {\em The electron bremsstrahlung spectrum from neutral atoms and
  ions},
\newblock In: C. F. Barnett and M. F. Harrison (eds.), in Applied Collision
  Physics. Academic Press, NY, 1984.

\bibitem{bjorken:65}
J.~D. Bjorken and S.~D. Drell,
\newblock {\em Relativistic Quantum Fields},
\newblock McGraw-Hill, NY, 1965.

\bibitem{itzykson:80}
C.~Itzykson and J.~{B}ernard Zuber,
\newblock {\em Quantum Field Theory},
\newblock Mc{G}raw-Hill, NY, 1980.

\bibitem{rose:61}
M.~E. Rose,
\newblock {\em {Relativistic Electron Theory}},
\newblock John Wiley, NY, 1961.

\bibitem{eichler:95:book}
J.~Eichler and W.~Meyerhof,
\newblock {\em Relativistic Atomic Collisions},
\newblock Academic Press, San Diego, 1995.
\newblock Note the misprinted overall sign in Eq.~(4.115).

\bibitem{varshalovich}
D.~A. Varshalovich, A.~N. Moskalev, and V.~K. Khersonski\u{i},
\newblock {\em Quantum Theory of Angular Momentum},
\newblock World Scientific, Singapure, 1988.

\bibitem{artemyev:03:ncdr}
A.~N. Artemyev, T.~Beier, J.~Eichler, A.~E. Klasnikov, C.~Kozhuharov, V.~M.
  Shabaev, T.~St\"ohlker, and V.~A. Yerokhin,
\newblock Phys. Rev. A {\bf 67}, 052711 (2003).

\end{thebibliography}

\end{document}